\documentclass[useAMS, twocolumn]{mn2e}
\usepackage{graphicx}
\usepackage{amssymb}
\usepackage{amsmath}

\newcommand\be{\begin{equation}}
\newcommand\ee{\end{equation}}

\newcommand\araa{ARA\&A}
\newcommand\apj{ApJ}
\newcommand\apjl{ApJL}
\newcommand\apjs{ApJS}
\newcommand\aap{A\&A}

\newcommand\icarus{Icarus}
\newcommand\mnras{MNRAS}
\newcommand\nat{Nature}
\newcommand\ssr{Space~Sci.~Rev.}

\title[Planet Traps in Truncated Discs]{Trapping of Low-Mass Planets Outside the Truncated Inner Edges of Protoplanetary Discs}
\author[R.~Miranda and D.~Lai]{Ryan Miranda\thanks{rjm456@cornell.edu} and Dong Lai \\
          Cornell Center for Astrophysics and Planetary Science, Department of Astronomy, Cornell University, Ithaca, NY 14853, USA}

\begin{document}

\maketitle

\begin{abstract}
We investigate the migration of a low-mass ($\lesssim 10 M_\oplus$) planet near the inner edge of a protoplanetary disc using two-dimensional viscous hydrodynamics simulations. We employ an inner boundary condition representing the truncation of the disc at the stellar corotation radius. As described by Tsang (2011), wave reflection at the inner disc boundary modifies the Type I migration torque on the planet, allowing migration to be halted before the planet reaches the inner edge of the disc. For low-viscosity discs ($\alpha \lesssim 10^{-3}$), planets may be trapped with semi-major axes as large as $3-5$ times the inner disc radius. In general, planets are trapped closer to the inner edge as either the planet mass or the disc viscosity parameter $\alpha$ increases, and farther from the inner edge as the disc thickness is increased. This planet trapping mechanism may impact the formation and migration history of close-in compact multiplanet systems.
\end{abstract}


\section{Introduction}

Tidal interactions of planets with their natal discs lead to angular momentum exchange and orbital migration (Goldreich \& Tremaine 1979, 1980; Lin \& Papaloizou 1979; Ward 1986, 1997). The direction of the orbital migration is usually inwards, although this depends on the structure, thermodynamics, and radiative and turbulent properties of the disc (see Kley \& Nelson 2012; Baruteau et al.~2014). For low-mass planets, with insufficient mass to modify the surface density profile of the disc, the resulting ``Type I'' migration is rapid, and proceeds until the planet reaches the inner edge of the disc or the magnetospheric cavity.

In Type I migration theory, the disc responds strongly to the gravitational potential of the planet at Lindblad resonances (LRs), at which the forcing frequency from the planet on the disc (as measured by a comoving fluid element) matches the radial epicyclic frequency of the disc. Waves excited at either outer Lindblad resonances (OLRs), which carry positive angular momentum, or at inner Lindblad resonances (ILRs), which carry negative angular momentum, propagate away from the planet until they are damped either viscously (e.g., Takeuchi et al.~1996) or non-linearly (e.g., Goodman \& Rafikov 2001). Migration is a result of the differential Lindblad torque (i.e., the asymmetry between the negative torque applied to the planet by OLRs and the positive torque by ILRs) and the corotation torque. In the standard theory, the disc is considered to be infinite in extent, so that waves travel exclusively away from the planet and damp far from it. In numerical studies of planet-disc interaction (e.g., de Val-Borro et al.~2006), this is achieved by imposing wave damping zones at the edges of the computational domain, so that all waves propagate away from the planet, and wave reflection from the edges of the domain is prohibited.

Tsang (2011; hereafter T11) suggested that if the inner edge of the disc is partially reflective, waves launched by the planet at ILRs may be trapped and form standing waves between the disc edge and the ILR, enhancing their positive torques on the planet. He used linear theory to calculate the ILR torque associated with trapped waves, and showed that the enhanced torque leads to the existence of migration traps, at which the migration of the planet is halted, with a semi-major axis several times larger than the radius of the disc inner edge. The limitation of linear theory prevents a quantitative determination of the location of migration traps, as several issues (such as the effects of wave damping and corotation torque) could not be adequately resolved (e.g., Kley \& Nelson 2012).

Protoplanetary discs may either extend to the stellar surface, or be truncated by the stellar magnetic field. In the latter case, accretion on to the central star is complex and three-dimensional inside of the truncation radius, involving interchange instabilities and the funneling of plasma along the stellar magnetic field lines (see, e.g., Lai 2014; Romanova et al.~2014 for recent reviews). In either case, the inner edge of the disc (i.e., disc-star or disc-magnetosphere interface) may present a semi-rigid ``wall'' to incoming waves (e.g., Tsang \& Lai 2009; Fu \& Lai 2012), allowing them to be reflected, and making the migration halting mechanism described by T11 possible.

In this paper, we numerically investigate the migration of a low-mass planet ($\lesssim 10 M_\oplus$) near the inner edge of a disc, using two-dimensional, viscous hydrodynamics simulations. We assume that the disc is truncated at the stellar corotation radius, in a configuration such that no angular momentum is transferred to the star. We show that Type I migration can indeed be halted at several (up to $5.4$ in our simulation survey) times the inner disc radius, and we explore how the locations of the migration traps depend on the planet mass and disc parameters (viscosity and thickness).

\section{Problem Setup}

We consider a planet with mass $M_\mathrm{p}$ and semi-major axis $a_\mathrm{p}$ embedded in a gaseous disc orbiting a star of mass $M_*$. The planet-to-star mass ratio is is $q = M_\mathrm{p}/M_*$. The disc is described in (cylindrical) polar coordinates $(r,\phi)$ by the surface density $\Sigma$ and velocity $(u_r, u_\phi)$. The disc is geometrically thin, and rotates at approximately the local Keplerian rate $\Omega_\mathrm{K} = (GM_*/r^3)^{1/2}$. Its radial extent is $r_\mathrm{in}$ to $r_\mathrm{out} = 7 r_\mathrm{in}$, and the orbital period at $r_\mathrm{in}$, $P_\mathrm{in} = 2\pi/\Omega_\mathrm{K}(r_\mathrm{in})$, is taken as the reference unit of time. The equation of state is locally isothermal, with $p = \Sigma c_\mathrm{s}^2(r)$, where $p$ is the height-integrated pressure, and
\be
c_\mathrm{s}(r) = h r \Omega_\mathrm{K}
\ee
is the sound speed. The disc aspect ratio, $h = H/r$, is chosen to have the constant value $h = 0.05$, unless otherwise specified. The kinematic viscosity is described by the $\alpha$ prescription,
\be
\nu = \alpha c_\mathrm{s} H,
\ee
and we choose several different values of $\alpha$ in this study.

\subsection{Disc Structure}

The disc is initialized in an equilibrium state, which is not strongly modified by the presence of the (low-mass) planet (i.e., the disc and planet are in the Type I migration regime). We adopt a state with a constant mass accretion rate, $\dot{M} = -2\pi r\Sigma u_r$. The disc structure is determined by conservation of angular momentum, which takes the form
\be
\dot{M}l + 2 \pi r^3 \nu \Sigma \frac{\mathrm{d}\Omega}{\mathrm{d}r} = \dot{M} l_0,
\ee
where the first term on the left-hand side represents inward advection of angular momentum, and the second term represents outward transport of angular momentum due to viscous stress. The quantity $l_0$ on the right-hand side is a constant, whose value is equal to the ratio of the angular momentum accretion rate and mass accretion rate through the disc (and on to the central star), $\dot{J}/\dot{M}$ (e.g., Popham \& Narayan 1991). The profiles of surface density and radial velocity satisfying this equation are
\be
\label{eq:sigma_1}
\Sigma(r) = \frac{\dot{M}}{3\pi\nu} \left(1 - \frac{l_0}{l}\right),
\ee
\be
\label{eq:ur_1}
u_r(r) = -\frac{3}{2}\frac{\nu}{r} \left(1 - \frac{l_0}{l}\right)^{-1}.
\ee
Additionally, the angular velocity is approximately Keplerian, but modified by the pressure gradient,
\be
u_\phi(r) = \left(v_\mathrm{K}^2 + \frac{r}{\Sigma}\frac{\mathrm{d}p}{\mathrm{d}r}\right)^{1/2},
\ee
where $v_\mathrm{K} = r\Omega_\mathrm{K}$ is the Keplerian orbital velocity.

\subsubsection{Inner Boundary}

The eigenvalue $l_0$ of the stationary disc is determined by the physics of the inner disc boundary. For non-magnetic stars, the disc stops at the stellar surface, $r_\mathrm{in} = R_*$. More likely, for magnetic stars, the disc is truncated at the magnetosphere radius, at which the magnetic stress due to the (dipolar) stellar field balances the plasma stress:
\be
r_\mathrm{m} = k\left(\frac{B_*^4 R_*^{12}}{GM_*\dot{M}^2}\right)^{1/7},
\ee
where $B_*$ is the stellar surface magnetic field strength and $k \approx 1$ is a constant. In both cases, to avoid complications associated with boundary layers, we assume that the flow inside inside $r_\mathrm{in}$ rotates at $\Omega_\mathrm{K}(r_\mathrm{in})$, i.e., $r_\mathrm{in}$ coincides with the corotation radius of the star,
\be
r_\mathrm{co} = \left(\frac{G M_*}{\Omega_*^2}\right)^{1/3}.
\ee
This represents the spin equilibrium of the star. The plausibility of such a configuration is supported by analytic and numerical studies of the interactions of discs with magnetic stars (e.g., Ghosh \& Lamb 1979; Koenigl 1991; Ostriker \& Shu 1995; Long et al.~2005). In this state, there is no net angular momentum transfer to the star, i.e., $l_0 = 0$. Equations~(\ref{eq:sigma_1}) and (\ref{eq:ur_1}) then become
\be
\label{eq:sigma_2}
\Sigma(r) = \Sigma_0 \left(\frac{r}{r_\mathrm{in}}\right)^{-1/2},
\ee
where $\Sigma_0 = \dot{M}/[3\pi\nu(r_\mathrm{in})]$, and
\be
\label{eq:ur_2}
u_r(r) = -\frac{3}{2}\alpha h^2 v_\mathrm{K}.
\ee

In general, fluid perturbations around $r_\mathrm{in}$ are complicated, and depend on the dynamics of the flow inside $r_\mathrm{in}$ (see Tsang \& Lai 2009; Fu \& Lai 2012). For simplicity, we adopt the following ``fixed'' boundary condition: we assume that at $r_\mathrm{in}$, the orbital velocity is fixed at its Keplerian value, while the surface density and radial velocity are fixed at their equilibrium values (i.e., equations~\ref{eq:sigma_2} and \ref{eq:ur_2} evaluated at $r_\mathrm{in}$):
\be
\label{eq:fixed_bc}
\left(\Sigma, u_r, u_\phi\right)_{r_\mathrm{in}} = \left[\Sigma_0, \left(-\frac{3\nu}{2r}\right)_{r_\mathrm{in}}, v_\mathrm{K}(r_\mathrm{in})\right].
\ee

\subsubsection{Disc Mass and Planet Migration}

To set the disc mass and migration time-scale, we first consider the standard Type I migration (i.e., with no wave reflection at the inner disc boundary). The total (Lindblad $+$ corotation) torque on the planet can be written as $\Gamma_\mathrm{tot} = -C\Gamma_0$, where
\be
\label{eq:gamma_0}
\Gamma_0 = \left(\frac{M_\mathrm{p}}{M_*}\right)^2 \left(\frac{H}{a_\mathrm{p}}\right)^{-2} \Sigma_\mathrm{p} a_\mathrm{p}^4 \Omega_\mathrm{p}^2
\ee
is the characteristic Type I migration torque [here $\Sigma_\mathrm{p} = \Sigma(a_\mathrm{p})$], and the order unity coefficient $C$ depends on the structure of the disc. For an infinite, two dimensional disc with $\Sigma \propto r^{-1/2}$ and $c_\mathrm{s} \propto r^{-1/2}$, $C = 1.47$ (D'Angelo \& Lubow 2010). The orbital evolution of the planet is determined by $\Gamma_\mathrm{tot} = \mathrm{d}{L}_\mathrm{p}/\mathrm{d}t$, where $L_\mathrm{p}$ is the angular momentum of the planet. Thus the semi-major axis decays exponentially, with a time-scale given by
\be
\label{eq:t_mig}
\begin{aligned}
t_\mathrm{mig} & = \frac{h^2}{4\pi Cq} \left(\frac{\Sigma_0 r_\mathrm{in}^2}{M_*}\right)^{-1} P_\mathrm{in} \\
& = 1.35 \times 10^4 \left(\frac{q}{10^{-5}}\right)^{-1} \left(\frac{h}{0.05}\right)^{2} \left(\frac{\Sigma_0 r_\mathrm{in}^2}{10^{-3} M_*}\right)^{-1} P_\mathrm{in}.
\end{aligned}
\ee
With the ``fixed'' inner disc boundary condition (equation~\ref{eq:fixed_bc}), the migration time-scale can be significantly modified from equation~(\ref{eq:t_mig}), due to the effect of wave reflection.

Throughout this paper, we choose $\Sigma_0 = 10^{-3} M_*/r_\mathrm{in}^2$, and have scaled equation~(\ref{eq:t_mig}) accordingly. Note that this value of $\Sigma_0$ is about $20$ times larger than the minimum-mass solar nebula profile (Hayashi 1981), $6.0 \times 10^{-5} M_*/r_\mathrm{in}^2$ (taking $r_\mathrm{in} = 0.1$ AU), which would lead to a longer migration time-scale ($2.26 \times 10^5 P_\mathrm{in}$) for our computation. However, it is only about two times larger than the ``minimum-mass extrasolar nebula'' profile of Chiang \& Laughlin (2013).

For most of our simulations, the initial semi-major axis of the planet is $a_0 = a_\mathrm{p}(t = 0) = 3 r_\mathrm{in}$. This is the approximate upper limit for the semi-major axis of migration traps based on the analytical estimate of T11. However, we find that the migration halting mechanism is somewhat more robust than indicated by the analytical results. In some cases (e.g., thick discs), the planet migrates negligibly when initialized at $3r_\mathrm{in}$, indicating that the migration trap lies beyond $3r_\mathrm{in}$, and we have used a much larger $a_0$ to determine the trap radius.

\subsection{Wave Damping Zones}

Near the outer disc boundary ($r/r_\mathrm{in} = 6.4 - 7.0$), we apply the wave damping conditions (e.g., de Val-Borro et al.~2006), in which the fluid variables are damped to their initial (unperturbed) values on the local orbital time-scale. This prevents wave reflection at the outer boundary, so that only outgoing waves may propagate (physically, this represents the existence of an extended disc beyond the outer boundary). Although mass and angular momentum are not conserved in the damping zone, this would only lead to inconsistent results over the time-scale required for the planet to to alter the structure of the outer disc (a few times $10^5 P_\mathrm{in}$; see the last paragraph of Section \ref{subsec:torques}), and so is not an issue in our simulations, which last $\sim 10^4 P_\mathrm{in}$. Near the inner boundary, we do not apply the damping conditions, as we are explicitly interested in the reflection of waves by the inner edge. However, for comparison, we have also performed several runs in which wave damping near the inner boundary ($r/r_\mathrm{in} = 1.0 - 1.3$) is included, to mimic the standard numerical planet-disc interaction setup, in which the disc extends to much smaller radii.

\subsection{Numerical Method}

The two-dimensional viscous hydrodynamic equations of the disc and the dynamics of the planet are solved using the code \textsc{fargo3d} (Ben{\'{\i}}tez-Llambay \& Masset 2016). The gravitational potential of the planet is softened over a length of $\epsilon = 0.6 H$. The shift of resonances due to the fact that the planet feels the gravity of the disc, while the disc does not, is corrected using the method described in Baruteau \& Masset (2008). Unless otherwise stated, the dimensions of the numerical grid are $N_r \times N_\phi = 477 \times 1536$ (see Section \ref{subsec:resolution} for a resolution study), with logarithmic spacing in $r$ and uniform spacing in $\phi$, so that the grid cells are approximately square ($\Delta r \approx r \Delta \phi$) everywhere. The planet is smoothly introduced to the disc by ramping its mass up from zero over the first $50 P_\mathrm{in}$ (approximately $10$ orbits of the planet at its initial $a_\mathrm{p} = 3 r_\mathrm{in}$).

\section{Results}

\subsection{Migration and Trapping}

\begin{figure}
\begin{center}
\includegraphics[width=0.49\textwidth,clip]{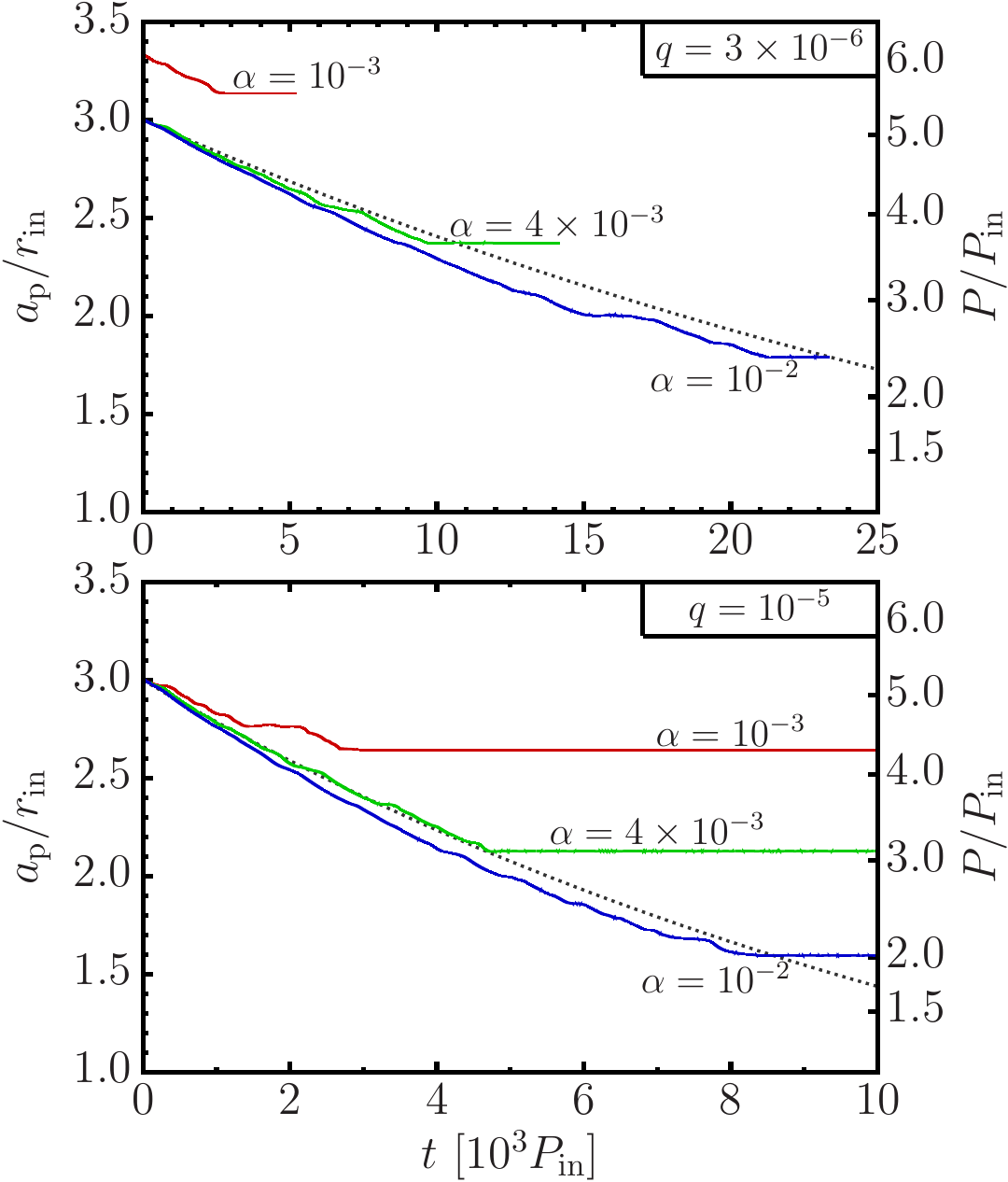}
\caption{The semi-major axis of the planet as function of time for $q = M_\mathrm{p}/M_* = 3 \times 10^{-6}$ (top panel) and $q = 10^{-5}$ (bottom panel), with different values of the viscosity parameter, $\alpha$ (solid lines). The disc aspect ratio is fixed at $h = 0.05$. The dotted lines correspond to the migration of the planet in an infinite disc. Note that the scale of the $x$-axis differs between the two panels, and that in the top panel, for the case with $\alpha = 10^{-3}$, the planet is initialized with a slightly larger semi-major axis than in the other cases.}
\label{fig:migration}
\end{center}
\end{figure}

\begin{figure}
\begin{center}
\includegraphics[width=0.40\textwidth,clip]{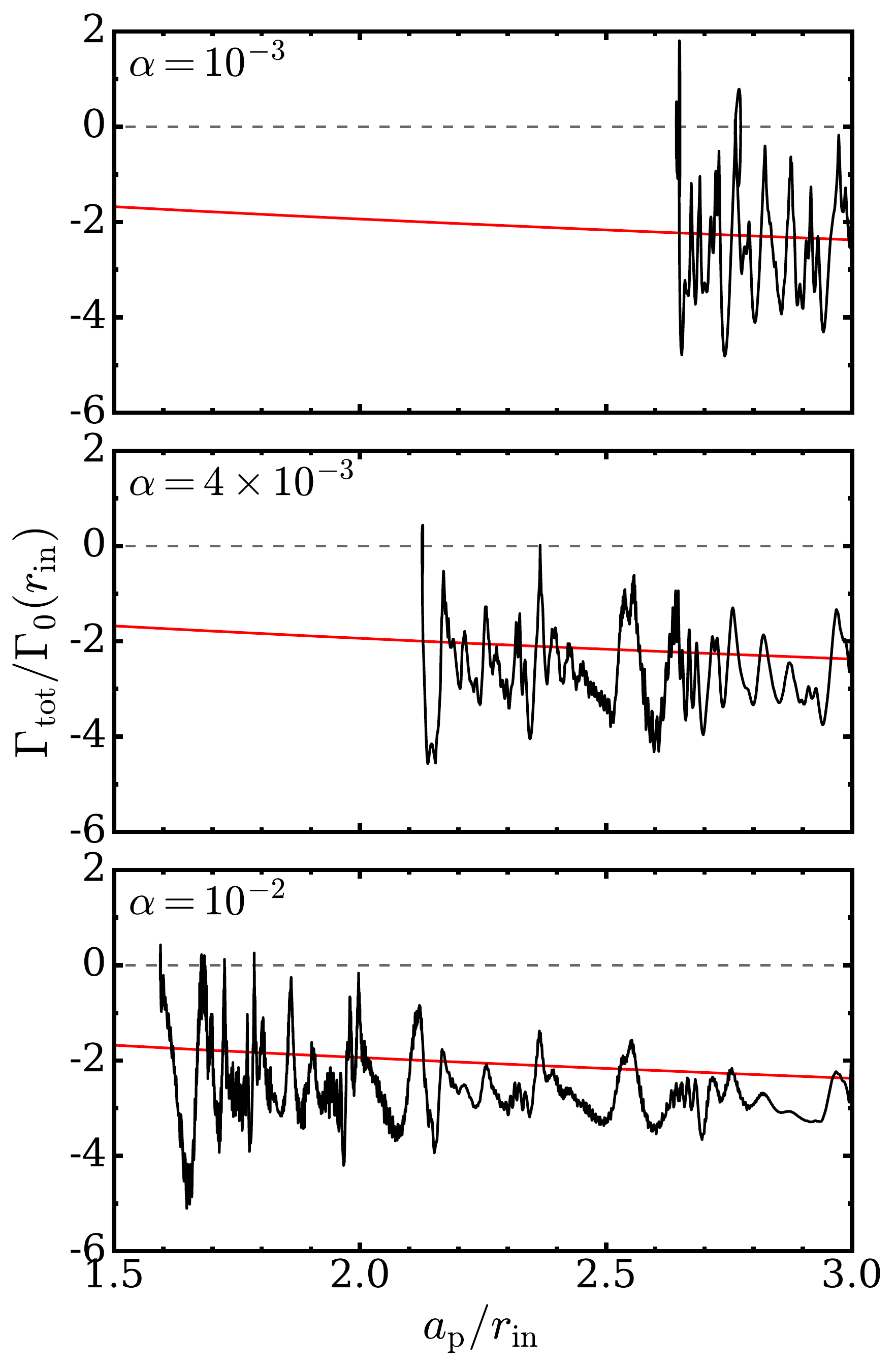}
\caption{Total torque (black curves) on the planet for the case with $q = 10^{-5}$, as a function of semi-major axis, for three different values of $\alpha$. The planet moves from right to left in these plots. The torque has been time-averaged over intervals of $5 P_\mathrm{in}$ in order to filter out short-time-scale variations. The red line represents the torque on a planet undergoing normal Type I migration. The location in the disc at which the planet ultimately stops migrating is coincident with a zero-crossing of $\Gamma_\mathrm{tot}$.}
\label{fig:torque_map}
\end{center}
\end{figure}

\begin{figure}
\begin{center}
\includegraphics[width=0.49\textwidth,clip]{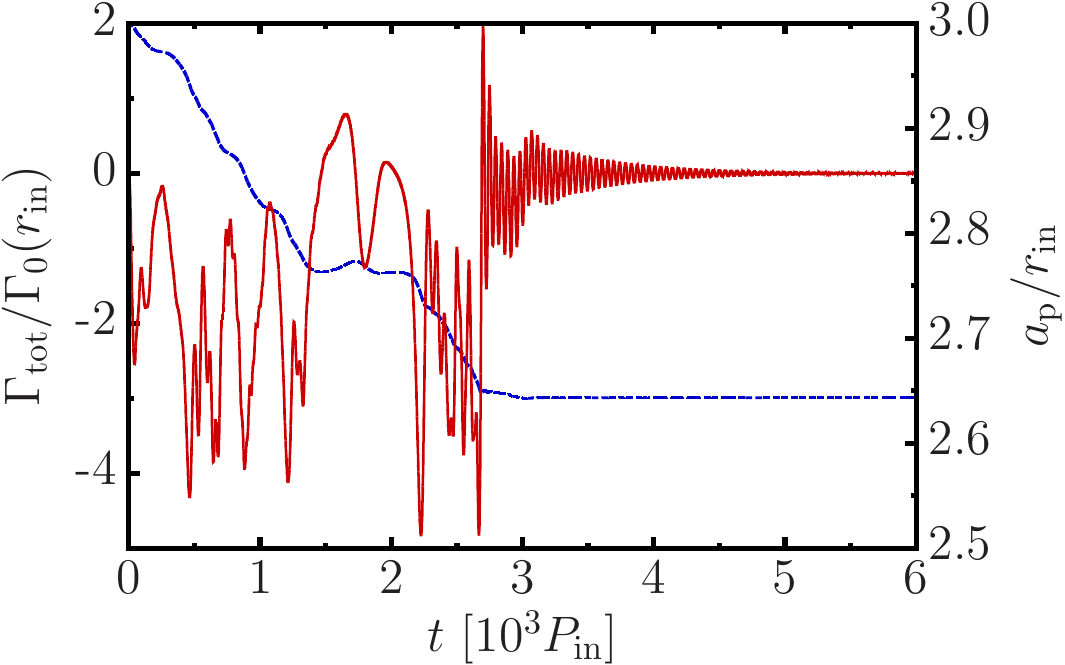}
\caption{Total torque on the planet (solid line, left axis) and semi-major axis (dashed line, right axis) as a function of time for the case with $q = 10^{-5}$ and $\alpha = 10^{-3}$.}
\label{fig:torque_t}
\end{center}
\end{figure}

\begin{figure}
\begin{center}
\includegraphics[width=0.49\textwidth,clip]{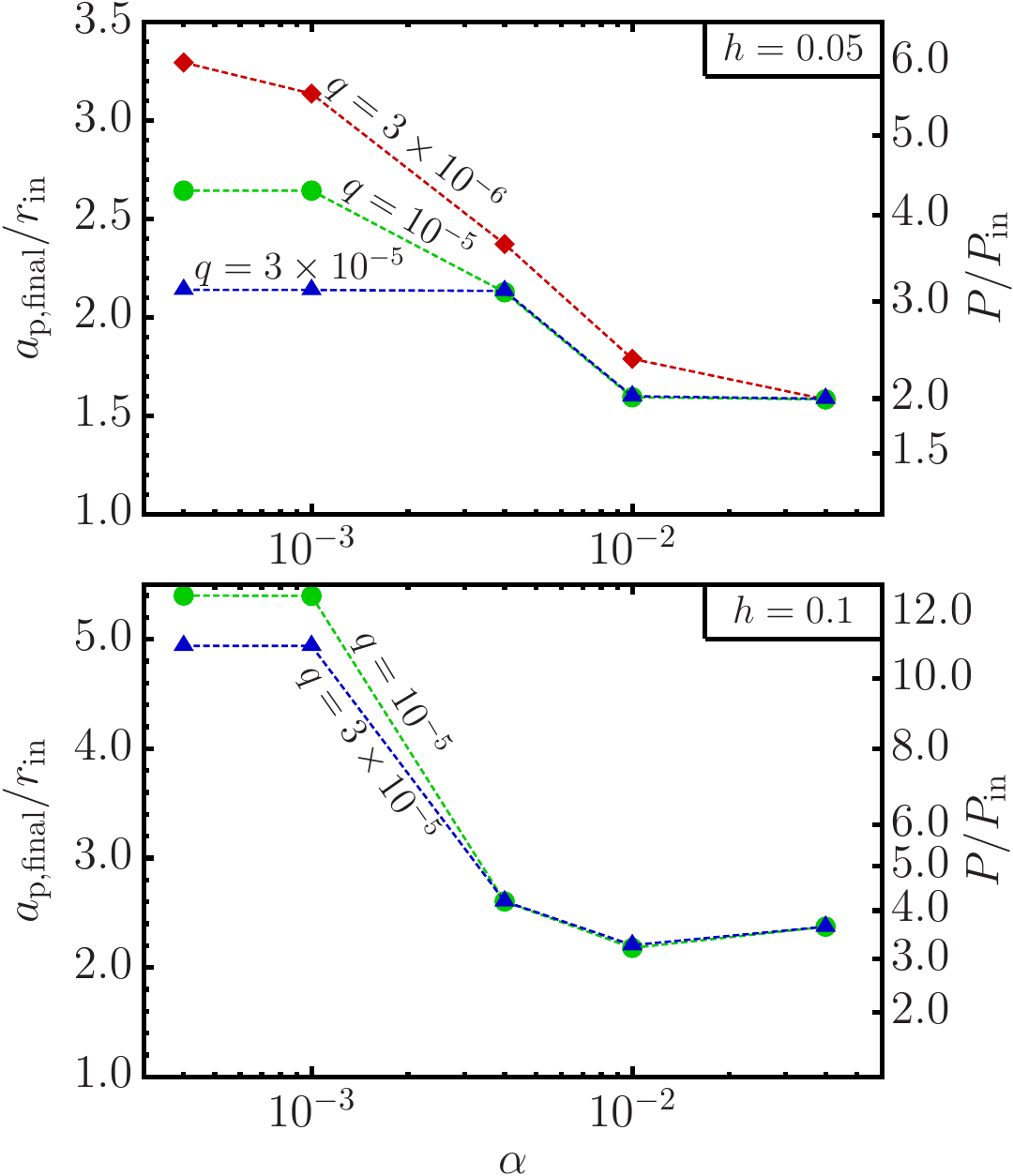}
\caption{The final semi-major axis of the planet, $a_\mathrm{p,final}$, as a function of $\alpha$, for different values of $q = M_\mathrm{p}/M_*$. In the top panel, the disc aspect ratio is $h = 0.05$ (the same value used in all other figures in this paper), while in the bottom panel, we adopt the larger value $h = 0.1$. The right axis label shows the orbital period of the planet in units of the period at $r_\mathrm{in}$.}
\label{fig:a_final}
\end{center}
\end{figure}

The migration (semi-major axis as a function of time) of the planet for several different mass ratios in discs with different values of the viscous $\alpha$ parameter is shown in Fig.~\ref{fig:migration}. The planet initially migrates inwards at a rate approximately equal to the migration rate in an infinite disc (equation~\ref{eq:t_mig}), on average. However, the instantaneous migration rate varies relative to the average (by a factor of order unity) on time-scales of hundreds to thousands times $P_\mathrm{in}$ (see Figs.~\ref{fig:torque_map}--\ref{fig:torque_t}). Eventually, the planet encounters a ``trap'' and stops migrating, with the semi-major axis remaining constant for the remainder of the simulation (at least several thousand $P_\mathrm{in}$ for all cases).

The total torque on the planet as a function of semi-major axis is shown in Fig.~\ref{fig:torque_map}, for the cases with $q = 10^{-5}$ and several values of $\alpha$ (cf. Fig.~3 of T11). The standard Type I migration torque is also shown for comparison. The torque is an oscillatory function of $a_\mathrm{p}$; averaged over a range of $a_\mathrm{p}$, the torque is approximately equal to the the standard torque, but it can also be significantly larger or smaller (in magnitude) at different locations in the disc. In particular, as described in T11, the torque has zero-crossings, corresponding to locations at which migration is halted. Figure~\ref{fig:torque_t} shows an example of the torque as a function of time, which decays towards zero as the planet settles into the trap. The analytical calculations presented in T11 suggest that the spatial variations in torque and presence of zero-crossings are largely suppressed for $\alpha = 10^{-3}$ (for a disc thickness $h = 0.05$). However, we find that these features are present for values of $\alpha$ as large as $0.04$.

In some cases, there are zero crossings which the planet migrates through without stopping. These correspond to intervals of time, for example $t \approx (1500 - 2500) P_\mathrm{in}$ for the case of $q = 10^{-5}$ and $\alpha = 10^{-3}$, for which migration is dramatically slowed down, but not completely stopped (see Fig.~\ref{fig:torque_t}). At these points, the planet spends an extended amount of time with a nearly constant semi-major axis, but eventually breaks out of the ``temporary'' migration trap and continues to migrate inwards, until ultimately encountering a ``permanent'' trap. This behaviour can be attributed to the short-term variations in the torque on the orbital period and grid cell crossing time of the planet (the latter is a numerical effect, see Section \ref{subsec:resolution}). If the amplitude of the torque variations is larger than the slowly-varying torque associated with a migration trap, then the planet can in principle escape from a trap.

The final semi-major axis of the planet, $a_\mathrm{p,final}$, is shown in Fig.~\ref{fig:a_final} (top panel), for planets with $q = 3 \times 10^{-6}$, $10^{-5}$, and $3 \times 10^{-5}$ (i.e., $M_\oplus$, $3 M_\oplus$, and $10 M_\oplus$, if $M_* = 1 M_\odot$), with a viscous $\alpha$ in the range $4 \times 10^{-4} - 4 \times 10^{-2}$. Generally, $a_\mathrm{p,final}$ decreases as either $q$ or $\alpha$ increases. For each $q$, there is a maximum $a_\mathrm{p,final}$, whose value decreases with $q$, which is achieved when $\alpha$ is sufficiently small ($\lesssim 10^{-3}$). As $\alpha$ is increased from $10^{-3}$ to $10^{-2}$, $a_\mathrm{p,final}$ decreases. For $\alpha \gtrsim 10^{-2}$, $a_\mathrm{p,final}$ is approximately constant, at a value of about $1.59 r_\mathrm{in}$. Although this is very close to the location of the $2$:$1$ orbital commensurability with the inner edge of the disc, it has no physical significance and is merely a numerical coincidence, since we find that $a_\mathrm{p,final}$ can be shifted away from this value with small changes in the value of $h$.

Also shown in Fig.~\ref{fig:a_final} (bottom panel) is $a_\mathrm{p,final}$ as a function of $\alpha$, for the case with a larger disc aspect ratio, $h = 0.1$. The results are qualitatively similar to the case with $h = 0.05$; $a_\mathrm{p,final}$ decreases with $\alpha$ before reaching a nearly constant minimum value for $\alpha \gtrsim 10^{-2}$ (there is a slight increase in $a_\mathrm{p,final}$ when $\alpha$ is further increased from this value, although its magnitude is small relative to the total range of $a_\mathrm{p,final}$). However, $a_\mathrm{p,final}$ is systematically larger compared to the case with smaller $h$, and is never smaller than $2.2 r_\mathrm{in}$. For $\alpha \lesssim 10^{-3}$, $a_\mathrm{p,final}$ can be as large as $5.4 r_\mathrm{in}$ (note that for these cases, the planet was initialized with a semi-major axis of $5.5 r_\mathrm{in}$, and the outer boundary of the numerical grid was extended to $12 r_\mathrm{in}$).

\subsection{Wave Reflection and Enhanced Inner Lindblad Torques}
\label{subsec:torques}

\begin{figure*}
\begin{center}
\includegraphics[width=0.99\textwidth,clip]{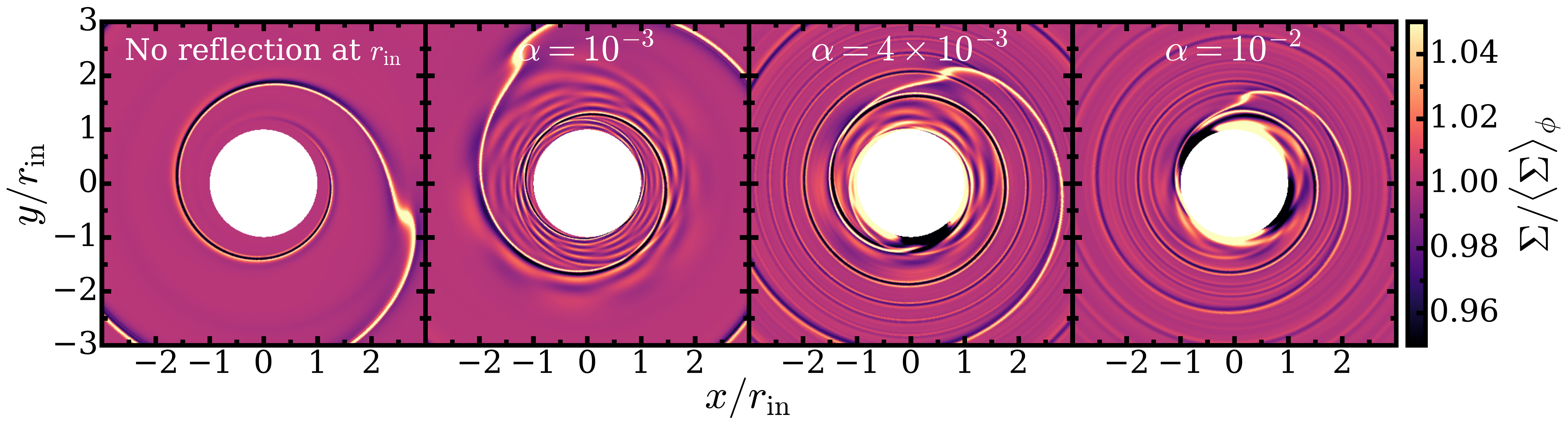}
\caption{Snapshots of the disc surface density, relative to the azimuthally-averaged surface density, for a planet with $q = 10^{-5}$. In the leftmost panel, the planet is undergoing normal Type I migration (i.e., an inner wave damping zone is included and there is no wave reflection at the inner disc edge). In the other panels, the planet is self-consistently trapped at its final semi-major axis, for different values of $\alpha$.}
\label{fig:snapshots}
\end{center}
\end{figure*}

\begin{figure}
\begin{center}
\includegraphics[width=0.45\textwidth,clip]{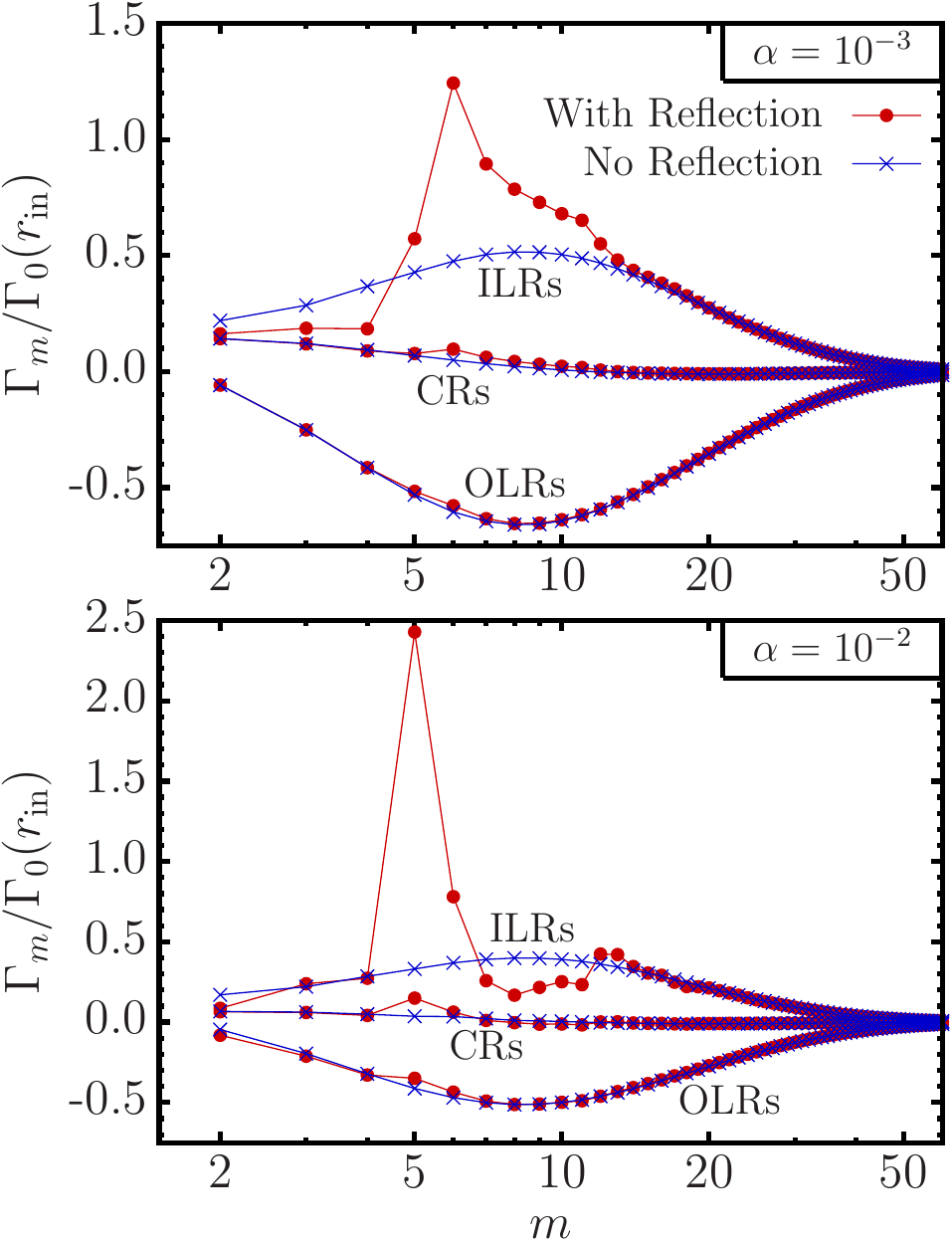}
\caption{The torque on the planet decomposed into components with different azimuthal numbers, $\Gamma_m$, and further separated into the contributions from corotation resonances (``CRs'') and inner/outer Lindblad resonances (``ILRs/OLRs''), which are operationally defined as originating from within the radial interval $a_\mathrm{p} \pm r_\mathrm{H}$, or interior/exterior to this region, respectively. Two cases are shown: the case in which wave reflection at the inner edge is allowed and the planet is self-consistently held in place by a resonant trap (filled circles), and the case in which wave reflection is prohibited by imposing an inner damping condition, and the semi-major axis of the planet is artificially held fixed (crosses). Here $q = 10^{-5}$, and the top and bottom panels correspond to two different values of $\alpha$. For the case with reflection, the amplitudes of the OLR and CR torques are largely unchanged compared to the case without reflection, but the amplitudes of the ILR torques are enhanced for a range of values of $m$, specifically $m = 5- 12$ for $\alpha = 10^{-3}$ and $m = 5-6$ for $\alpha = 10^{-2}$.}
\label{fig:torque_m}
\end{center}
\end{figure}

To examine the origin of the migration trap, Fig.~\ref{fig:snapshots} shows snapshots of the disc surface density (relative to the local azimuthally-averaged surface density), when the planet is at its final semi-major axis, for the case with $q = 10^{-5}$ and several different values of $\alpha$. Also shown for comparison is the case of a freely migrating planet, with an inner wave damping zone included so that wave reflection at the inner boundary is prohibited. For the cases in which the planet is trapped, there are visible standing waves between the inner edge and the orbit of the planet.
 
We decompose the torque on the planet into components which have different azimuthal numbers and which originate from different parts of the disc, according to
\be
\Gamma_m = \pi \int r \Sigma_m(r) \left(\frac{\partial \Phi}{\partial \phi}\right)_m(r) \mathrm{d}r,
\ee
where $\Sigma_m$ and $(\partial\Phi/\partial \phi)_m$ are the $m$-components (i.e., parts proportional to $e^{\mathrm{i}m\phi}$) of the surface density and (softened) tidal potential. The integral is taken over either $[r_\mathrm{in}, a_\mathrm{p} - r_\mathrm{H}]$ for torques due to ILRs, $[a_\mathrm{p} - r_\mathrm{H}, a_\mathrm{p} + r_\mathrm{H}]$ for corotation resonances (CRs), or $[a_\mathrm{p} + r_\mathrm{H}, r_\mathrm{out}]$ for OLRs. Here $r_\mathrm{H} = (q/3)^{1/3} a_\mathrm{p}$ is the Hill radius of the planet. Note that these are convenient operational definitions, rather than exact definitions, for the torques due to the different types of resonances. The decomposition is shown in Fig.~\ref{fig:torque_m}, for the cases with $q = 10^{-5}$ and two different values of $\alpha$. Both the case of a (self-consistently) trapped planet, and the case of a planet fixed at the same location, with an inner wave damping zone included (so that wave reflection is prohibited), are shown. The torques due to OLRs and CRs are mostly unaffected by reflection, while the ILR torques are enhanced for a range of $m$ values. For $\alpha = 10^{-3}$, the enhanced ILRs have $m = 5-12$, and for $\alpha = 10^{-2}$, they have $m = 5-6$. Torque components with small values of $m$ are preferentially enhanced when $\alpha$ is large, since the ILRs with large values of $m$ are located too far from the inner edge, and waves launched there tend to be damped before reaching the inner edge.

In addition to the reflection effect, a surface density enhancement of the inner disc could also lead to enhanced ILR torques. However, we find that there is minimal evolution of the azimuthally-averaged surface density profile in our simulations when the fixed boundary condition is used; variations from the initial profile are at the level of a few per cent. A surface density enhancement could occur due to the modification of the disc structure by the planet, on a time-scale $t_\mathrm{evolve} \sim L_\mathrm{D}/\Gamma_\mathrm{ILRs}$, where $L_\mathrm{D} \sim \Sigma r_\mathrm{in}^4 \Omega_\mathrm{in}$ is the inner disc angular momentum and $\Gamma_\mathrm{ILRs} \sim h^{-1}\Gamma_0$ is the sum of the one-sided Lindblad torques. For typical parameters (e.g., $q = 10^{-5}$, $\alpha = 0.01$), $t_\mathrm{evolve} \sim 1.4 \times 10^5 P_\mathrm{in}$, which is much longer than the duration of our simulations. It can also be longer than the inner disc viscous time-scale, $t_\mathrm{visc} \sim r_\mathrm{in}^2/\nu(r_\mathrm{in}) \sim 6 \times 10^3 P_\mathrm{in}$ (for the same parameters), which may indicate that the disc can be maintained in its viscous equilibrium state despite the influence of the planet. Since the mass of the inner disc changes negligibly in our simulations, this effect does not contribute to the modification of the ILR torques.

\subsection{Alternative Inner Boundary Condition}
\label{subsec:bc}

\begin{figure}
\begin{center}
\includegraphics[width=0.49\textwidth,clip]{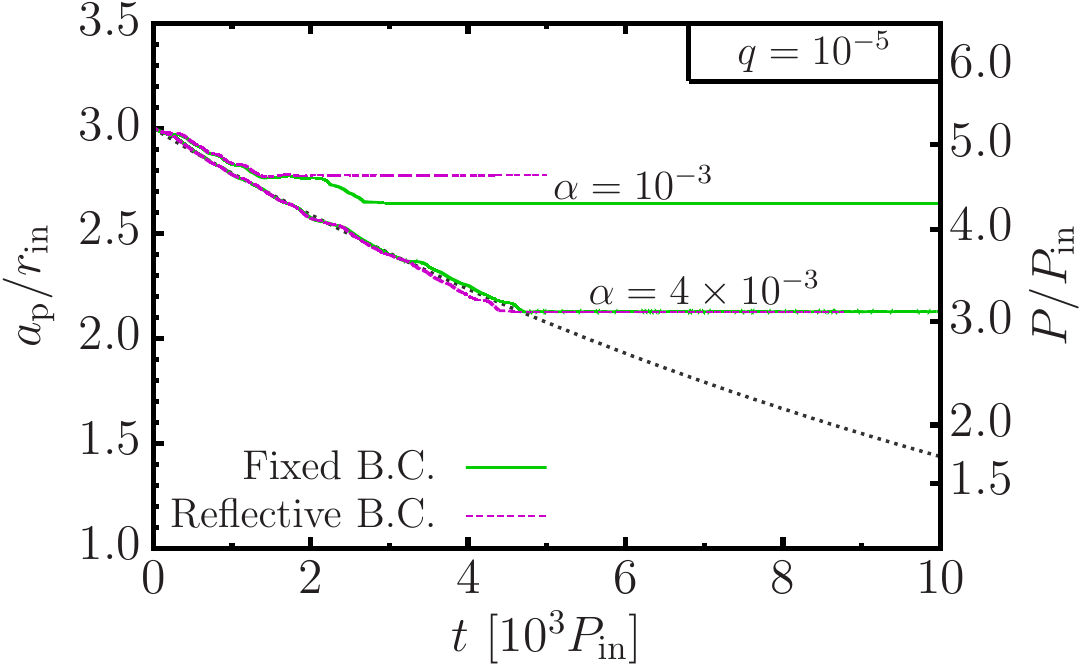}
\caption{The semi-major axis as a function of time for $q = 10^{-5}$, and with different values of $\alpha$, using two different boundary conditions: the standard ``fixed'' boundary condition (equation~\ref{eq:fixed_bc}), and the ``reflective'' boundary condition (equation~\ref{eq:reflective_bc}). Neither the migration history nor the final semi-major axis of the planet are strongly affected by the particular choice of the inner boundary condition.}
\label{fig:migration_bc}
\end{center}
\end{figure}

The ``fixed'' inner boundary conditions we have applied (equation~\ref{eq:fixed_bc}), in which all fluid variables are held at fixed values, represent a highly simplified model of the disc-magnetosphere boundary. To assess the robustness of the planet trapping phenomenon found above, we have investigated an alternative boundary condition, namely a ``reflective'' boundary condition, in which $\Sigma$ and $u_\phi$ are symmetrized across the boundary and $u_r$ is antisymmetrized, so that
\be
\label{eq:reflective_bc}
\left(\frac{\partial \Sigma}{\partial r}, u_r, \frac{\partial u_\phi}{\partial r}, \right)_{r_\mathrm{in}} = (0, 0, 0).
\ee
The results of this numerical experiment, for the cases with $q = 10^{-5}$ and two different values of $\alpha$, are shown in Fig.~\ref{fig:migration_bc}. The migration history of the planet is similar, but not identical to the case with the ``fixed'' boundary conditions, and ultimately, the migration halts at a nearly the same location. For $\alpha = 10^{-3}$, migration halts at $2.78 r_\mathrm{in}$, which is the location of a ``temporary trap'' at which the planet temporarily stalls in our fixed boundary condition simulation. This represents a $5$ per cent difference in the final semi-major axis between the two different boundary conditions. For $\alpha = 4 \times 10^{-3}$, the planet stops at $2.13 r_\mathrm{in}$ regardless of the choice of the two inner boundary conditions. We note that, unlike the ``fixed'' boundary condition, the ``reflective'' boundary condition leads to significant mass accumulation in the inner disc (as a result of the inward viscous drift of the gas and its inability to penetrate the inner boundary, where $u_r = 0$). However, this evidently does not significantly affect the trapping of the planet. We conclude that our results are not sensitive to the details of the inner boundary condition as long as a sharp truncated edge is present.

\subsection{Numerical Resolution}
\label{subsec:resolution}

There is a resolution-dependent numerical effect which may influence the migration and trapping of the planet. The torque on the planet varies as it migrates through the numerical grid, with a period equal to the radial grid cell crossing time, $|\Delta r/\dot{a}_\mathrm{p}|$, where $\Delta r$ is the grid spacing (Masset \& Papaloizou 2003), and with an amplitude which decreases as the resolution increases. To test how numerical resolution affects the final semi-major axis of the planet, we have carried out a resolution study for the case with $q = 10^{-5}$ and $\alpha = 10^{-3}$. Our standard numerical resolution is $N_r \times N_\phi = 477 \times 1536$, so we also ran this case using a lower resolution ($318 \times 1024$) and using a higher resolution ($636 \times 2048$). The final semi-major axis $a_\mathrm{p,final}$ decreases by $3$ per cent (compared to the standard resolution) for the low-resolution run, and increases by $0.5$ per cent for the high-resolution run. The numerical results are therefore converged with respect to resolution.

\section{Discussion}

\subsection{Dependence on Viscosity}
\label{subsec:viscosity}

The planet is trapped closer to the inner edge of the disc as the viscosity parameter $\alpha$ is increased (see Fig.~\ref{fig:a_final}). This is in qualitative agreement with the results of T11, in which viscous damping reduces the maximum possible enhancement of ILR torques responsible for stopping the migration of the planet. However, the final semi-major axis is relatively insensitive to $\alpha$ for $0.01 \lesssim \alpha \lesssim 0.04$.

T11 estimates the critical viscosity parameter, $\alpha_\mathrm{crit}$, above which Type I migration can no longer be halted by trapped modes, by considering the effect of viscosity on the resonant gain associated with an individual $m$-mode, and finds that $\alpha_\mathrm{crit} \sim h^{5/2} \sim 6 \times 10^{-4}$.  Numerically, we find that trapping is possible even for the largest value of $\alpha$ used in our experiments ($0.04$). We attribute this discrepancy to the fact that modes with multiple values of $m$ can be simultaneously resonant when the planet is trapped (see Fig.~\ref{fig:torque_m}), and so $\alpha$ can be much larger than the case of a single mode. This is because, if the ILR torque of a single $m$-mode were to balance out all of the OLR torques, then it would have to be enhanced by a factof of $\sim h^{-1}$ (approximately the total number of modes contributiong to the total torque), and so a small amount of viscous damping would make this impossible. However, if the ILR torques associated with many values of $m$ collectively cancel out the OLRs, then they must each only be slightly enhanced, and so the cancellation does not depend as sensitively on $\alpha$

The lowest $\alpha$ we have considered is $4 \times 10^{-4}$. For discs with lower viscosities, i.e., ``nearly laminar'' discs, Type I migration is modified due to the opening of a partial gap and density feedback effects, which may slow down or stop migration (Hourigan \& Ward 1984; Ward \& Hourigan 1989; Ward 1997; Rafikov 2002; Li et al.~2009; Fung \& Chiang 2017). We find that when $\alpha \lesssim 10^{-3}$, the trapping radius tends to approach a constant value (independent of $\alpha$).

\subsection{Dependence on Planet Mass}

The semi-major axis at which the planet is trapped decreases as the mass of the planet increases. As suggested by T11, the trapping mechanism is modified when the planet is massive enough for non-linear damping (due to shocks) to play a role. In particular, when $q \gtrsim (2/3)h^3 \sim 8 \times 10^{-5}$, non-linear effects are dominant, since waves excited by the planet shock immediately (Goodman \& Rafikov 2001). This may set an approximate upper limit for the mass ratios of planets which may be trapped. Indeed, in one of our numerical experiments (not shown), a planet with $q = 10^{-4}$ proceeds to migrate nearly all the way to the inner edge of the disc. In addition, a sufficiently massive planet may open a gap in the disc, which greatly slows its migration (Type II migration). This occurs when (Crida et al.~2006)
\be
1.08 h q^{-1/3} + 50 \alpha h^2 q^{-1} \lesssim 1,
\ee
i.e., when $q \gtrsim (10^{-4} - 10^{-3})$, depending on the value of $\alpha$. Therefore, there may only exist a narrow range of planet masses which result in rapid (Type I) migration and proceeds all the way to the inner edge of the disc.

\subsection{Dependence on Disc Thickness}

Throughout most of this paper, we adopted a disc thickness of $h = H/r = 0.05$, and found that planets can be trapped at about $(1.6-3.3) r_\mathrm{in}$. Additional numerical experiments presented in Fig.~\ref{fig:a_final} show that for a thicker disc with $h = 0.1$, the final semi-major axis of the planet is at least $2.2 r_\mathrm{in}$, and can be as large as $5.4 r_\mathrm{in}$ for small $\alpha$. This is significantly larger than the maximum possible trapping radius of about $3 r_\mathrm{in}$ predicted by T11. The fact that trapping is effective farther from the inner edge for thicker discs can be understood as follows. The resonance condition which determines the locations of possible planet traps requires that a small integer number of wavelengths fit between the ILR and the inner edge of the disc. Since the wavelength of density waves (for a given wave frequency) increases with $h$, the trapping condition can be satisfied at locations farther from $r_\mathrm{in}$ for thicker discs.

\subsection{Caveats}

In this paper we have implemented a simple inner disc boundary condition to mimic the disc-magnetosphere boundary, which in reality is a complicated, three-dimensional structure. Our simulations can therefore not be compared to state-of-the-art simulations of planet migration (including three-dimensional effects, magnetic fields, radiative transfer, etc.) or star-disc interactions (see Romanova \& Owocki 2015). Rather, they serve to highlight the significant effect of a partially reflective inner disc edge on the migration of a planet, which has until now been neglected in numerical studies. Our numerical experiments with two very different inner boundary conditions (``fixed'' and``reflective''; see equations~\ref{eq:fixed_bc} and \ref{eq:reflective_bc}) suggest that planet trapping due to a sharp, truncated disc inner edge is a robust phenomenon. Nevertheless, it remains to be seen whether or not this effect persists in more realistic simulations which include a more complete treatment of the relevant physics.

The simulations presented in this paper are two-dimensional and employ a locally isothermal sound speed profile and parametrized viscosity. In three dimensions, shocks resulting from the vertical deflection of waves due to refraction (Lin et al.~1990) or channeling of wave energy towards the disc surface (Lubow \& Ogilvie 1998; Ogilvie \& Lubow 1999) could enhance the effective viscosity of the disc and modify the properties of trapped modes. We do not include the effects of magnetic fields, turbulence, or self-gravity, all of which may modify planet migration (see Kley \& Nelson 2012; Baruteau et al.~2014). In particular, turbulence can significantly alter, and in some cases inhibit Type I migration (e.g., Nelson \& Papaloizou 2004). Therefore, it could in principle disrupt the trapping mechanism explored in this paper.

Density fluctuations associated with ideal MRI turbulence lead to stochastic torques with an rms amplitude  
\be
\Gamma_\mathrm{turb} \sim \alpha^{1/2} M_\mathrm{p} \frac{\Sigma_\mathrm{p} a_\mathrm{p}^2}{M_*} a_\mathrm{p}^2 \Omega_\mathrm{p}^2
\ee
(e.g., Johnson et al.~2006; Okuzumi \& Ormel 2013). A planet in a migration trap is stably maintained in the trap by a positive torque that it experiences if it moves inward, whose magnitude is $\sim \Gamma_0$ (equation~\ref{eq:gamma_0}; see Figs.~\ref{fig:torque_map}--\ref{fig:torque_t}). The planet can therefore stay in the trap only if $\Gamma_\mathrm{turb}/\Gamma_0 \lesssim 1$, i.e., if
\be
\label{eq:turbulence_condition}
\alpha^{1/2} \left(\frac{H}{r}\right)^2 \lesssim q.
\ee
This is also the approximate criterion which determines if the planet can undergo smooth Type I migration rather than diffusive turbulent migration. For the parameter values we have considered ($q \sim 10^{-5}$, $h \sim 0.05$), a turbulence level of $\alpha \lesssim 10^{-5}$ is required for the planet to stay in a trap. This places the greatest restriction on the disc properties necessary for the trapping mechanism to be effective. However, turbulence in protoplanetary discs can be significantly suppressed by non-ideal MHD effects (e.g., Bai \& Stone 2013; Turner et al.~2014; Lesur et al.~2014; Simon et al.~2015), so the $\alpha$ parameter in equation~(\ref{eq:turbulence_condition}) that characterizes the strength of turbulence can be very small.

\subsection{Observational Implications}

Disc-driven migration may have played an important role in the assembly of exoplanetary systems with multiple coplanar, transiting planets, such as Kepler 11 (Lissaeur et al.~2011) and TRAPPIST-1 (Gillon et al.~2017). The effect of magnetospheric truncation may be imprinted on the current architecture of such systems (Mulders et al.~2015). This could result from the truncation radius setting an innermost initial orbital radius for planets formed in-situ (Lee \& Chiang 2017), or from the migration of planets formed farther out (e.g., Liu et al.~2017). When migration occurs, our results indicate that, instead of migrating all the way to the disc edge, the planet may be halted at several (as large as $5.4$ in our simulations) times the radius of the inner edge. This could have important implications for the formation of these planetary systems. For example, it may be difficult to explain how planets with close-in orbits acquired these orbits, without invoking additional physical mechanisms to bring them inwards. It may also suggest that these planets must be formed or delivered to their current locations in the relatively early phase of the disc evolution.

\section{Conclusion}

We have carried out two-dimensional hydrodynamical simulations of a low-mass planet ($q = 3 \times 10^{-6} - 3 \times 10^{-5}$, or $M_\mathrm{p} = M_\oplus - 10 M_\oplus$ for a $1 M_*$ star) undergoing Type I migration in a disc with a distinct inner edge, such as the disc-magnetosphere boundary. We assume that the disc is truncated at the corotation radius of the central star, and adopt two different boundary conditions to mimic the partial reflection of the disc edge to incoming density waves. In qualitative agreement with the theory of Tsang 2011 (T11), we find that the migration of the planet can be halted at a semi-major axis several times larger than the inner disc radius $r_\mathrm{in}$. This occurs due to the resonant enhancement of inner Lindblad torques as a result of stationary waves trapped between the orbit of the planet and the inner edge of the disc. A range of azimuthal mode numbers (e.g., $m \approx 5 - 12$) can contribute to the trapping, although this range becomes narrower and tends towards smaller values for more viscous discs. The semi-major axis at which the planet is trapped can be as large as about $5.4 r_\mathrm{in}$ in our simulations, and generally decreases with increasing planet mass or viscosity parameter, and increases with increasing disc thickness (see Fig.~\ref{fig:a_final}). This planet trapping mechanism may play an important role in shaping the architecture of exoplanetary systems of short-period planets.

\section*{Acknowledgements}
We thank Pierre-Yves Longaretti and David Tsang for useful discussion and comments. This work has been supported in part by NASA grants NNX14AG94G and NNX14AP31G, NSF grant AST-1715246, and a Simons Fellowship from the Simons Foundation.

\end{document}